\documentclass[11pt,onecolumn]{revtex4}
\usepackage{
graphics,
amsmath,
amssymb,
rotating,
times,
alltt,
}
\newcommand{\suppress}[1]{}
\renewcommand{\marginpar}[1]{}

\begin{document}

\title{
Predicting Genetic Regulatory Response using Classification:\\ Yeast Stress Response}

\author{Manuel Middendorf, Anshul Kundaje,\\ 
Chris Wiggins, Yoav Freund, Christina Leslie}

\affiliation{Columbia University, New York NY 10027, USA}

\begin{abstract}
We present a novel classification-based algorithm
called GeneClass for learning to predict
gene regulatory response.
Our approach is motivated by the hypothesis that
in simple
organisms such as {\it Saccharomyces cerevisiae}, we can learn
a decision rule for predicting whether a gene is up- or
down-regulated in a particular experiment based on (1)
the presence of binding site
subsequences
(``motifs'')
in the gene's regulatory region
and (2) the expression levels of regulators such as
transcription factors in the experiment (``parents'').
Thus our learning task
integrates
two
qualitatively different data sources:
genome-wide cDNA microarray data
across multiple perturbation and mutant experiments along with
motif profile data from regulatory sequences.
Rather than focusing on the
regression task of
predicting real-valued gene expression measurements, GeneClass
performs the
classification task of predicting +1 and -1 labels, corresponding to
up- and down-regulation beyond the levels of biological and
measurement noise in microarray measurements.
GeneClass uses the Adaboost learning algorithm with a margin-based
generalization of decision trees called alternating decision trees.
In computational experiments based on the Gasch
{\it S. cerevisiae} dataset,
we show that the GeneClass method
predicts up- and down-regulation on held-out experiments with high
accuracy.
We explore a range of experimental setups related to environmental
stress response, and we retrieve important regulators, binding site
motifs, and relationships between regulators and binding sites that
are known to be associated to specific stress response pathways.
Our method thus provides predictive hypotheses, suggests
biological experiments, and provides interpretable insight into
the
structure
of genetic regulatory networks.
\end{abstract}

\maketitle

\noindent{\bf Supplementary website:}
\url{http://www.cs.columbia.edu/compbio/geneclass}

\section{Introduction}

Understanding the underlying mechanisms of gene
transcriptional regulation through
analysis of high-throughput genomic data
has become an important current research area in computational biology.
For simpler model organisms such as {\em S. cerevisiae},
there have been numerous computational approaches
that combine
gene expression
data from microarray experiments and regulatory sequence data
to solve
different problems in gene regulation:
identification of regulatory elements in non-coding DNA \cite{bussemaker:reduce,hughes:alignace},
discovery of co-occurrence of regulatory motifs
and combinatorial effects of regulatory molecules \cite{pilpel:motifsyn}, and organization of genes that appear to be
subject to common regulatory control into ``regulatory
modules'' \cite{ihmels:module,segal:module}.
Most of the recent
studies
can be placed broadly
in one of three categories: {\em statistical approaches}, which aim to
identify
statistically significant regulatory
patterns in a dataset
\cite{bussemaker:reduce,pilpel:motifsyn,ihmels:module}; {\em probabilistic approaches},
which
try to discover structure in the dataset as formalized by probabilistic models
(often graphical models or Bayesian networks) \cite{segal:module,segal:learningmod,hartemink:graphical,peer:inferring,peer:minreg}; and
{\em linear network models}, which hope to learn explicit parameterized
models for pieces of the regulatory network by fitting to data \cite{collins:svd,hassle:lin}.  While these
approaches provide useful exploratory tools
that allow the user to generate
biological hypotheses about transcriptional regulation,
in general, they
are not yet adequate
for making accurate {\em predictions} about which genes will be up- or
down-regulated in new or held-out experiments.
Since these approaches do not emphasize prediction accuracy, it is
difficult to directly compare performance of the different
algorithms or decide, based
on cross-validation experiments, which approach is most likely to generate
plausible biological hypotheses for testing in the lab.

In the current work, we present an algorithm called GeneClass that
learns a {\em prediction} function for the
regulatory response of genes under different experimental conditions.
The inputs to our learning algorithm are
the gene-specific regulatory sequences -- represented by the set of
 binding site patterns they contain (``motifs'') -- and
the experiment-specific expression levels of regulators (``parents'').
The output
is a prediction of the
expression state 
of
the regulated gene.
Rather than trying to predict a real-valued expression level,
we formulate the task as a binary classification problem, that is,
we predict only whether the gene is up- or down-regulated.
This reduction allows us to exploit modern and effective
classification algorithms.
GeneClass uses the Adaboost learning algorithm
with
a margin-based generalization of decision trees called
alternating decision trees (ADTs).  Boosting, like support vector
machines, is a large-margin classification algorithm that performs
well for high-dimensional problems.
We evaluate the performance of our method
by measuring prediction accuracy on
held-out microarray experiments, and we achieve very good classification
results in this setting.
Moreover, we can analyze the
learned prediction trees to extract significant features or relationships
between features that
are associated with accurate generalization rather than just
correlations in the training data.
In a range of computational experiments for the investigation
of environmental stress response in yeast, GeneClass retrieves
significant regulators, binding motifs, and motif-regulatory pairs
that are known to be associated with specific stress response
pathways.

Among recent statistical approaches, the most
revelant method related to GeneClass
is the REDUCE algorithm of Bussemaker {\em et al.}~\cite{bussemaker:reduce} for regulatory
element discovery.  Given
gene expression measurements from a single microarray experiment and the
regulatory sequence $S_g$ for each gene $g$ represented on the array,
REDUCE proposes a linear model for the dependence of log gene expression
$E_g$
on presence of regulatory
subsequences
(or ``motifs")
\newcommand{\M}{\mu}
$E_g = C + \sum_{\M \in S_g} F_\M N_{\M g}$, where $N_{\M g}$ is a count of occurrences of regulatory
subsequence
$\M$ in sequence $S_g$, and the $F_\M$ are experiment-specific fit parameters.
GeneClass generalizes beyond the conditions of a single experiment
by using paired (motif$_g$,parent$_e$) features, where the parent variable
represents over- or under-expression of a regulator (transcription factor, signaling molecule, or protein kinase) in the experiment $e$, rather than
using motif information alone.
Note, however, that GeneClass uses classification rather
than regression as in REDUCE.

Restriction to a candidate set of potential parents
has also been used
in the probabilistic model literature, including
in the regression-based work of
Segal {\em et al.}
for partitioning target genes into {\em regulatory modules} for
{\em S. cerevisiae} \cite{segal:module}.  Here, each module is
a probabilistic
regression tree, where internal nodes of
the tree correspond to states of regulators
and each
leaf node prescribes a normal distribution
describing the expression of all the module's genes
given the regulator conditions.  The authors provide
some 
validation on new experiments by
establishing that the target gene sets of specific modules
do have statistically significant overlap with the set of
differentially expressed genes;
however,
they do not focus on making accurate predictions of differential
expression as we
do here.
Our GeneClass method retains the distinction between regulator
(``parent'') genes and target (``child'') genes, as well as a
model that can capture combinatorial relationships among
regulators; however, the margin-based GeneClass trees are very
different from probabilistic trees. Unlike in \cite{segal:module},
we learn from both expression and sequence data, so that the
influence of a regulator is mediated through the presence of a
regulatory sequence element.  We note that in separate work, Segal
{\em et al.} \cite{segal:learningmod} present a probabilistic
model for combining promoter sequence data and a large amount of
expression data to learn transcriptional modules on a genome-wide
level in {\em S.\ cerevisiae}, but they do not demonstrate how to
use this learned model for predictions of regulatory response.

The current work follows up on our original paper introducing
the GeneClass algorithm for prediction of regulatory
response \cite{middendorf:geneclass}.
Here, we report additional
computational experiments and more detailed biological validation
for specific environmental
stress responses (Section \ref{sec:biovalidation}).  Due to space
constraints, we omit some algorithmic details and
refer the reader to
the earlier presentation and to additional results
available at the supplementary website:
\url{http://www.cs.columbia.edu/compbio/geneclass}.

\section{Learning algorithm}
\subsection{Adaboost}

The underlying classification algorithm that we use
is Adaboost, introduced by Freund and
Schapire~\cite{Schapire02}, which works by
repeatedly applying a
simple learning algorithm, called the {\em weak} or {\em base} learner,
to different weightings of the same training set. For binary prediction
problems, Adaboost learns from a
training set that consists of pairs
$(x_1,y_1),(x_2,y_2),\ldots,(x_m,y_m)$, where $x_i$ corresponds to the
features of an example and $y_i \in \{-1,+1\}$ is the binary label
to be predicted, and maintains a {\em weighting} that
assigns a non-negative real value $w_i$ to each example
$(x_i,y_i)$.
On iteration $t$ of the boosting process, the weak learner is  applied
to the training set with weights $w_1^t,\ldots,w_m^t$ and
produces a prediction rule $h_t$ that maps $x$ to
$\{0,1\}$.
The rule
$h_t(x)$ is required to have a small but significant correlation
with the labels $y$  when measured using the
current weighting. After the function $h_t$ is
generated, the example weights are changed so that the weak
predictions $h_t(x)$ and the labels $y$ are decorrelated. The weak
learner is then called with the new weights over the training examples
and the process repeats.  Finally, one takes a linear combination
of all the weak prediction rules
to obtain a real-valued {\em strong} prediction
function or {\em prediction score} $F(x)$.
The strong prediction rule is given by
$\mbox{sign}(F(x))$:  
\begin{center}
\begin{tabular}{|l|}
\hline
 $F_0(x)\equiv0$\\
 for $t=1\ldots T$\\
\hspace{.5 cm} $w_i^t=\exp(-y_iF_{t-1}(x_i))$\\
\hspace{.5 cm} Get $h_t$ from {\it weak learner}\\
\hspace{.5 cm} $
\alpha_t=\ln\left(\frac
        {\sum_{i:h_t(x_i)=1, y_i=1} w_i^t}
        {\sum_{i:h_t(x_i)=1, y_i=-1} w_i^t}
\right) $\\
\hspace{.5 cm} $F_{t+1}=F_t+\alpha_th_t$.\\
\hline
\end{tabular}
\end{center}

One can prove that if the weak rules are
all slightly correlated
with the label, then the
strong rule learned by Adaboost will have a very high correlation
with the label -- in other
words, it will predict the label very accurately.
Moreover, one often observes that
the test
error of the strong rule (percentage of mistakes made on new examples)
continues to decrease even after the training error (fraction of
mistakes made on the training set) reaches zero.
This behavior has been related to the concept of a
``margin'', which is simply the value $y F(x)$~\cite{SchapireFrBaLe98}.
While $y F(x)>0$ corresponds to a correct prediction, $y F(x)>a>0$
corresponds to a {\em confident} correct prediction, and the
confidence increases monotonically with $a$.
Our experiments in this
paper demonstrate the correlation between large margins and correct
predictions on the test set
(see Section \ref{sec:experiments}).

\suppress{
\begin{figure}[t]
\begin{center}
\begin{tabular}{|l|}
\hline
 $F_0(x)\equiv0$\\
 for $t=1\ldots T$\\
\hspace{1 cm} $w_i^t=\exp(-y_iF_{t-1}(x_i))$\\
\hspace{1 cm} Get $h_t$ from {\it weak learner}\\
\hspace{1 cm} $
\alpha_t=\ln\left(\frac
        {\sum_{i:h_t(x_i)=1, y_i=1} w_i^t}
        {\sum_{i:h_t(x_i)=1, y_i=-1} w_i^t}
\right) $\\
\hspace{1 cm} $F_{t+1}=F_t+\alpha_th_t$\\
\hline
\end{tabular}
\end{center}
\caption{{\footnotesize {\bf The Adaboost algorithm.}} \label{fig:Adaboost}}
\end{figure}
}

\subsection{ADTs for Predicting Regulatory Response}
\renewcommand{\P}{\pi}

\suppress{
\begin{figure}[htb]
\begin{center}
\includegraphics[scale=0.25]{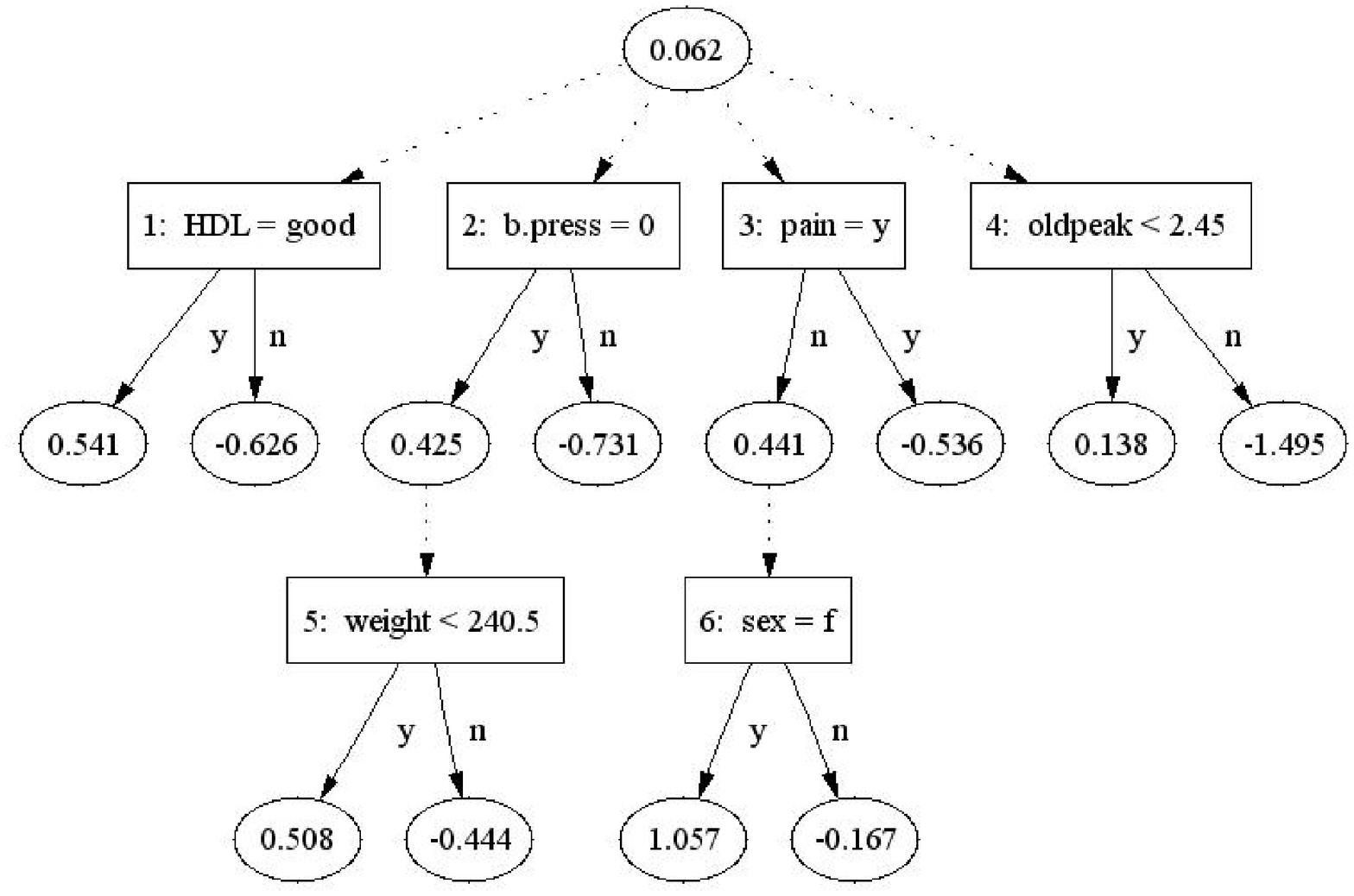}
\caption{\footnotesize {\bf An example ADT.}}\label{fig:adtexample}
\end{center}
\end{figure}
}

Adaboost is often used with a decision tree learning algorithm as
the base learning algorithm.
For the problem of predicting regulatory response,
we use a form of Adaboost
that produces a single tree-based decision rule
called an {\em alternating decision
tree} (ADT)~\cite{freund:alt}.  More details on learning ADTs for
regulatory response can be found in \cite{middendorf:geneclass}.

\begin{figure}[htb]
\begin{center}
\begin{tabular}{cc}
\resizebox{!}{1.9in}{\includegraphics{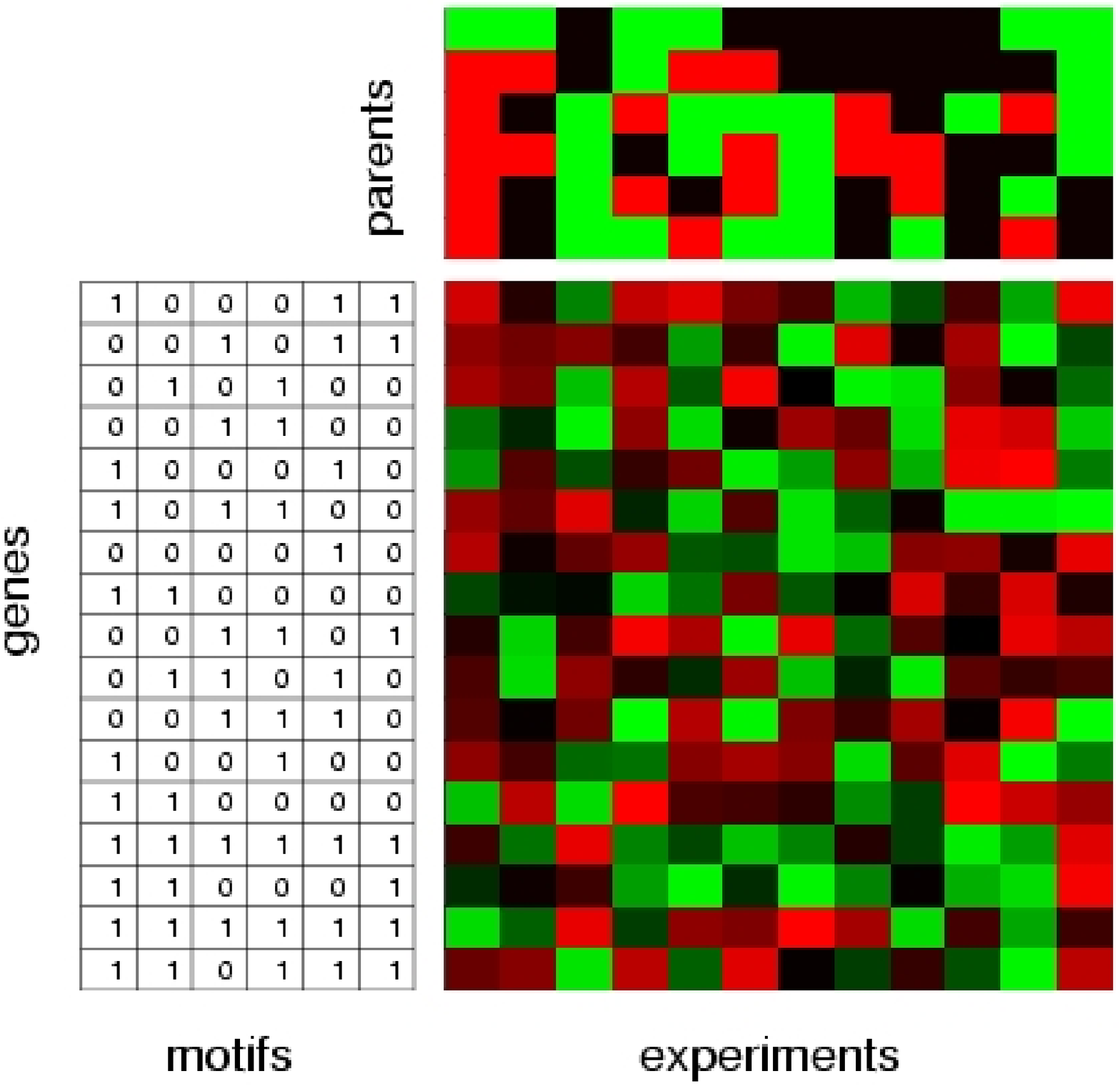}}~~ &
~~\resizebox{!}{1.9in}{\includegraphics{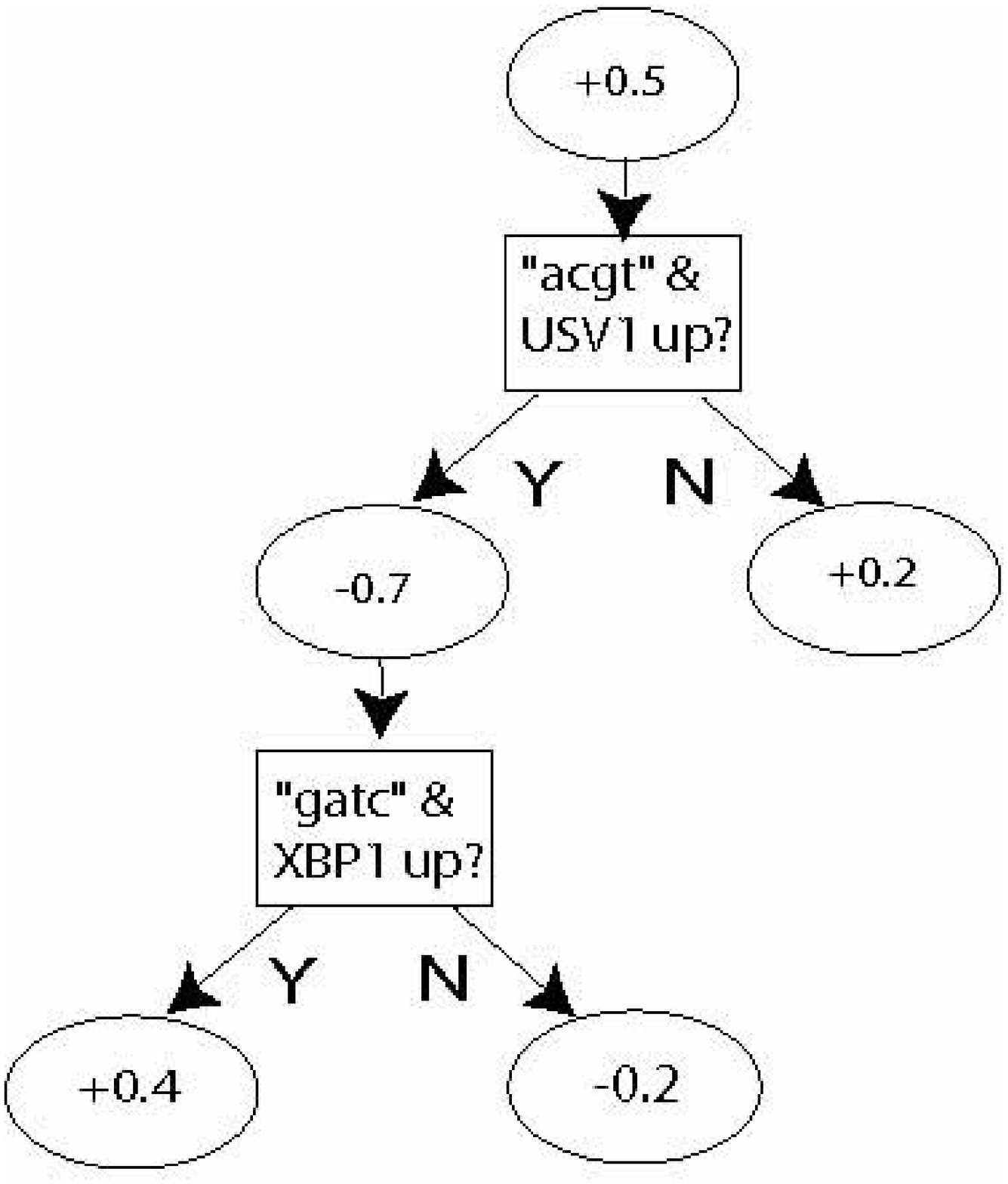}}\\
(A) & (B)
\end{tabular}
\end{center}
\caption{\footnotesize {\bf Boosting ADTs for regulatory 
response prediction.} 
In (A), we show the data presentation for our problem.
Every (target gene,experiment) is assigned a label of
+1 (up-regulated, in red) or -1 (down-regulated, in green) and
represented by the gene's vector of
motif counts (only binary values shown here) and the experiment's vector of regulator expression states.  A toy example of an ADT is shown in (B)
\label{fig:datarep}}
\end{figure}

Briefly, in our problem setting, we begin
with a candidate set of {\em motifs} $\M$ representing known or putative
regulatory element sequence patterns and a candidate set of regulators or
{\em parents} $\P$.
For each (gene,experiment) example in our gene expression dataset,
we have two sources of feature information relative to the candidate
motifs and candidate parent sets: a vector $N_{\M g}$ of motif counts of occurrences
of patterns $\M$ in the regulatory sequence of gene $g$,
and the vector $\P_e\in\{-1,0,1\}$
of expression states for parent genes $\P$ in the experiment $e$.
The data
representation is depicted in Figure \ref{fig:datarep} (A).

Figure \ref{fig:datarep} (B) shows a toy example of an ADT that could
be produced by Adaboost in our setting.  An ADT consists of alternating levels of {\em
prediction nodes} (ovals) -- which contain real-valued contributions to
the prediction scores -- and {\em splitter nodes} (rectangles) -- which
contain true/false conditions.  To obtain the
prediction score $F(x)$ for a particular
example $x$, we sum the values in all prediction nodes that we can reach
along {\em all} paths down from the root corresponding to yes/no decisions
consisent with $x$.

\suppress{
We explain the structure of ADTs using the example given in
Figure~\ref{fig:adtexample} reproduced from~\cite{freund:alt}.
The problem domain is heart disease diagnostics and the goal is to
predict whether an individual is healthy or sick based on 13 different
indicators.  The tree consists of alternating levels of ovals ({\em
prediction nodes}) and rectangles ({\em splitter nodes}).
The numbers within the ovals
define contributions to the prediction score. In this example,
positive contributions are evidence of a healthy heart, negative
contributions are evidence of a heart problem.  To evaluate the
prediction for a particular individual we start at the top oval
($0.062$) and follow the arrows down. We follow {\em all} of the
dotted arrows that emanate from prediction nodes, but we follow
{\em only one}
of the solid-line arrows emanating from a splitter node,
corresponding to the answer (yes or no) to the condition stated
in rectangle. We sum the values in all the prediction nodes
that we reach. This sum represents the prediction score $F(x)$ above,
and its sign is the prediction.

For example, suppose we had an individual for which
{\sc hdl=bad},
{\sc b.press=0}, {\sc pain=y},
{\sc oldpeak=2}, {\sc weight=300}, {\sc sex=m}.
In this case, the prediction
nodes that we reach in the tree are
$0.062,-0.626,0.425,-0.444,-0.536,0.138$, and summing 
gives a
score of $-0.981$, i.e., a very
confident diagnosis that the individual has a heart problem.

The ADT in the figure was generated by Adaboost from training data.
}

Splitter
nodes in our ADTs depend on decisions based on (motif,parent) pairs.
However, instead of splitting on real-valued thresholds of parent expression
and integer-valued motif count thresholds, we consider only whether a
motif $\M$ is present or not, and only whether a parent $\P$ is
over-expressed (or under-expressed) in the example.  Thus, splitter
nodes make boolean decisions based on conditions such as ``motif
$\M$ is
present and regulator $\P$ is over-expressed
(or under-expressed)''.
Paths in the learned ADT correspond to
conjunctions (AND operations) of these boolean (motif,parent) conditions.
Full details on selection of the candidate
motifs and regulators and discretization into up and down states is given
in Section \ref{sec:methods}.

In terms of Adaboost,
each prediction node represents a weak prediction rule, and at
every boosting iteration, a new splitter node together with its two
prediction nodes is introduced. The splitter node can be attached to
any previous prediction node, not only leaf nodes.
\suppress{
Each prediction
node is associated with a weight $\alpha$ that contributes to the
prediction score of every example reaching it.  The weak hypothesis
$h(x)$ is 1 for every example reaching the prediction node and 0 for
all others.
}
In general, more important decision rules are added at
early iterations.
We use this heuristic to
analyze the ADTs and identify the
most important factors in gene regulatory response.

\section{Methods}
\label{sec:methods}

\indent{\bf\em Dataset:} We use the Gasch {\em et al.}~\cite{gasch:yeast}
environmental stress response dataset,
consisting of
cDNA microarray experiments measuring genomic expression in
{\em S. cerevisiae} in response to diverse environmental
transitions.
There are a total of
6110 genes and 173 experiments
in the dataset,
with
all measurements given as $\log_2$ expression values (fold-change with
respect to
unstimulated reference expression).
We do not perform a zero mean
and unit variance normalization over experiments,
since we must retain the meaning of the true zero (no fold change).

{\bf\em Motif set:} We obtain the 500 bp 5' promoter sequences of
all \emph{S. cerevisiae} genes from the Saccharomyces Genome
Database (SGD). For each of these sequences, we search for
transcription factor (TF) binding sites using the PATCH software
licensed by TRANSFAC \cite{wigender:transfac}. The PATCH tool uses
a library of known and putative TF binding sites, some of which
are represented by position specific scoring matrices and some by
consensus patterns, from the TRANSFAC Professional database.
A total of 354 binding
sites
are used after pruning to remove redundant and rare sites.

{\bf\em Parent set:} We compile different sets of candidate
regulators to study the performance and dependence of our method
on the set of regulators. We restrict ourselves to a superset of
475 regulators (consisting of transcription factors, signaling
molecules and protein kinases), 466 of which are used in Segal
{\em et al.}
\cite{segal:module} and 9 generic (global) regulators obtained
from Lee {\em et al.}~\cite{lee:binding}.

Due to computational limitations on the number
of (motif,parent) features we could use in training, we select
smaller subsets of
regulators based on the following selection criteria. We use 13
high-variance regulators that had a standard deviation (in
expression over all experiments) above a cutoff of 1.2.
The second subset consists of the 9 global regulators that are
present in the Lee {\em et al.} studies but absent in the
candidate list of Segal {\em et al.} We also include 30 regulators
that are found to be top ranking regulators for the 50 clusters
identified in Segal {\em et al.} The union of these three lists
gives 53 unique regulators.

{\bf\em Target set and label assignment:} We discretize expression
values of all genes into three levels representing down-regulation
(-1), no change (0) and up-regulation (+1) using cutoffs based on
the empirical noise distribution around the baseline (0)
calculated from the three replicate unstimulated (time=0)
heat-shock experiments \cite{gasch:yeast}. We observe that 95\% of
the samples in this distribution had expression values between
$+1.3$ and $-1.3$. Thus we use these cutoffs to decide what we define
as significantly up-regulated (+1) and down-regulated (-1) beyond
the levels of biological and
experimental
noise in the microarray
measurements.

\suppress{
It is important to note that we train only on those
(gene,experiment) pairs for which we get a discretization of +1 or
-1, not examples where there is a baseline 0 label.
However, we are able to make predictions on {\em every} example in a
held-out experiment by thresholding on predicted margins, that is,
we abstain from predicting (predict baseline) if a prediction has margin
below threshold (see Section \ref{sec:experiments}).
}

Note that, although we {\it train} only on those
(gene,experiment) pairs which discretize to up- or down-regulated
states, we can {\it test} (make predictions) on every example
in a held-out experiment by thresholding on predicted margins.
That is, we \marginpar{abstaining from predicting is not
the same as predicting baseline-cw}
predict baseline if a prediction has margin
below threshold (see Section \ref{sec:experiments})).

We reduce our target gene list to a set of 1411 genes which
include 469 highly variant genes (standard deviation $>$ 1.2 in
expression over
all experiments) and 1250 genes that are part of the 17 clusters
identified by Gasch et al. \cite{gasch:yeast} using hierarchical
clustering (eliminating overlaps).

{\bf\em Software:} We use the MLJAVA software developed by Freund
and Schapire
\cite{schapire:boostexter}
which
implements the ADT learning algorithm. We use the text-feature in
MLJAVA to take advantage of the sparse motif matrix and use memory
more efficiently.

\section{Experimental Results}
\label{sec:experiments}

\subsection{Cross-Validation Experiments}
\label{sec:tenfoldcv}

We first perform cross-validation experiments to evaluate
classification performance on held-out
experiments. We divide
the set of 173 microarray experiments into 10 folds, grouping
replicate experiments together to avoid bias, and perform 10-fold
cross-validation experiments using boosting with ADTs on all 1411
target genes.

We train the ADTs for 400 boosting iterations, during most of
which
test-loss
decreases continuously.
We obtain an accuracy of 88.5\% on
up- and down-regulated
examples averaged over
10 folds (test loss of 11.5\%), showing that
predicting regulatory response is indeed possible in our framework.

To assess the difficulty of the classification task, we also compare
to a baseline method, $k$-nearest neighbor classification (kNN),
where each test example is classified by a vote of its $k$ nearest
neighbors in the training set.
For a distance function, we optimize
the
weighted sum of Euclidean distances
for motif and parent vectors, trying
values of $k <
20$ and both binary or integer representations of the
motif data (see \cite{middendorf:geneclass}). We obtain
minimum test-loss of 31.3\% at k=19 and with integer motif counts,
giving much poorer performance than boosting with ADTs.

Since ADTs output a real-valued prediction score $F(x)$
whose absolute value measures the
confidence of the classification, we can predict
a baseline label by thresholding on this score, that is,
we predict examples to be up- or down-regulated
if $F(x)>a$ or $F(x)<-a$ respectively, and to be baseline if
$|F(x)|<a$, where $a>0$.
\begin{figure}[htb]
\begin{center}
\begin{tabular}{cc}
\resizebox{!}{1.8in}{\includegraphics{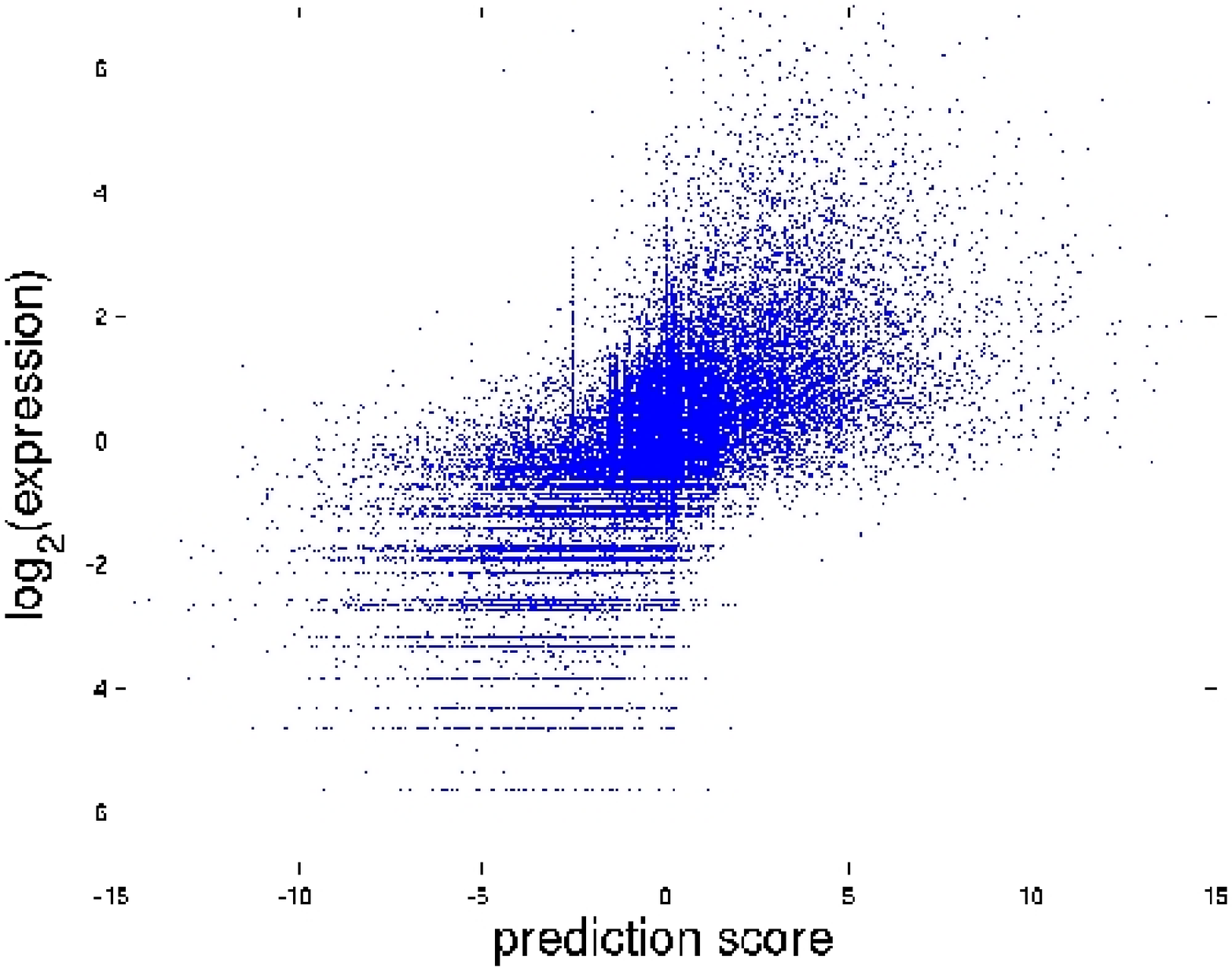}} &
\raisebox{.9in}{
\begin{normalsize}
\begin{tabular}{ll|ccc}
        &  & \multicolumn{3}{c}{Predicted Bins} \\
    &  & Down & Baseline & Up\\
\hline
           & Down          & 16.5\% & 8.9\% & 1.5\% \\
True Bins  & Baseline      & 9.3\% & 32.4\% & 6.3\% \\
           & Up            & 2.8\% & 9.9\% & 12.0\% \\

\end{tabular}
\end{normalsize}
}\\
(A) & (B) \\
\end{tabular}
\end{center}
\caption{\footnotesize {\bf True expression values versus prediction scores $F(x)$.}  The scatter plot (A) shows a high correlation between
prediction scores (x-axis) and true log expression values (y-axis) for genes on held-out experiments.  The confusion matrix (B) gives
truth and predictions for all genes in the held-out experiments,
including those expressed at baseline levels. Examples are binned by assigning a threshold $a = \pm0.5$ to
expression and prediction scores.
\label{figure:scatter}}
\end{figure}
Figure \ref{figure:scatter} (A)
shows expression values versus prediction scores for all
examples (up, down, and baseline) from the held-out experiments
using 10-fold cross-validation. The plot shows a high
correlation between expression and prediction, reminiscent of the
actual regression task.
Assigning thresholds to expression and
prediction values binning the examples into up, down and baseline
we obtain the confusion matrix in
Figure \ref{figure:scatter} (B).

\suppress{
\begin{table}
\begin{center}
\begin{scriptsize}
\begin{tabular}{ll|ccc}
        &  & \multicolumn{3}{c}{Predicted Bins} \\
    &  & Down & Baseline & Up\\
\hline
           & Down          & 16.5\% & 8.9\% & 1.5\% \\
True Bins  & Baseline      & 9.3\% & 32.4\% & 6.3\% \\
           & Up            & 2.8\% & 9.9\% & 12.0\% \\
\end{tabular}
\end{scriptsize}
\end{center}
\caption{\footnotesize {\bf Confusion Matrix:}
Truth and predictions for all genes in the held-out experiments,
including those expressed at baseline levels. Examples are binned by assigning a threshold $a = \pm0.5$ to
expression and prediction scores.
}
\label{table:confusion}
\end{table}
}

\subsection{Extracting features for biological interpretation}
\label{sec:biofeatures}
We describe below several approaches for extracting important features
from the learned ADT model, and we suggest ways to relate
these features to the biology of gene regulation.

{\bf\em Extracting significant features:}
We rank motifs, parents and motif-parent pairs by two main methods.
The {\em iteration score (IS)} of a feature is the  boosting
iteration during which it first appears in the ADT.
This ranking scheme appears to be useful in identifying
important motifs and motif-parent pairs
(restricting to iteration scores $< 50$), since features selected at early
rounds tend to be most significant.  The {\em abundance score (AS)}
of a regulator in the number of nodes in the final tree that include
the regulator as the parent in a motif-parent parent.
A regulator with a large abundance
score will affect a large number of paths through the ADT and
hence affect a large number of target genes.  If the state of a regulator
is changed through stress response or knockout, its predicted effect on
target genes will depend on its abundance in the ADT.
Note that presence of a strong motif-parent feature
does not necessarily imply a direct binding relationship between parent
and motif.
Such a pair could represent
an indirect regulatory relationship or some other kind of
predictive correlation, for example,
co-occurrence of the true binding site
with the motif corresponding to the feature.

{\bf\em ``In silico'' knock-outs:}
By removing a candidate from the regulator list and retraining the ADT,
 we can evaluate whether test loss significantly
decreases with omission of the parent and identify other weaker
regulators that are also correlated with the labels.  We investigate
in silico knock-outs in the biologically-motivated experiments described
in Section \ref{sec:biovalidation}

\subsection{Biological Validation Experiments}
\label{sec:biovalidation}
We designed the following four training and test sets of
selected microarray experiments to
study the response to specific types of stress.
By
comparative analysis of the trees learned from these
sets, we find and validate regulators
that are associated to regulation programs activated
by different stresses.
More detailed
results can be found on the supplementary website.

{\bf\em Heat-shock and osmolarity stress response:} In the first
study, we train on heat-shock, osmolarity, heat-shock knockouts,
over-expression, amino acid starvation experiments, and we test on
stationary phase, simultaneous heat-shock and hypo-osmolarity
experiments.

We observe a low test loss of 9.3\%, supporting the hypothesis
that pathways involved in heat-shock and osmolarity stress appear
to be independent
and the joint response to both stresses can be predicted easily.
This result agrees with the observation by Gasch {\em et al.}
\cite{gasch:yeast} that these two environmental stresses have
nearly additive effects on gene expression of environmental stress
response (ESR) genes. The high test accuracy also supports the
observation by Gasch {\em et al.} \cite{gasch:yeast} that the
response as well as parts of the underlying regulatory mechanisms
for stationary phase induction (test set) are similar to that of
heat-shock (training set).

The top five high scoring parents (based on AS) were USV1, XBP1, PPT1, GIS1 and
TPK1. These regulators are known or believed to play specific
important roles in each of the training and test set stresses.
Segal {\em et al.} \cite{segal:module} specifically identify USV1
as an important regulator in stationary phase (test set) and PPT1
to be important in the response to osmolarity stress (training
set).

The top ranking motif (based on IS) was the STRE
element of MSN2/MSN4, a known regulatory element
for a significant number of general stress response
genes \cite{gasch:yeast}.
The
connection of the osmolarity response to the HOG and other 
MAP kinase pathways is well known. Also, the osmolarity response
is strongly related to glycerol metabolism and transport
and hence closely associated with gluconeogenesis and glucose
metabolism pathways. We find the binding sites of CAT8
(gluconeogensis), GAL4 (galactose metabolism), MIG1 (glucose
metabolism and regulator of osmosensitive glycerol uptake)
\cite{klein:mig1}, GCN4 (regulator of HOG pathway and amino acid
metabolism), HSF1 (heat-shock factor), CHA4 (amino acid
catabolism), MET31 (methionine biosynthesis) and RAP1 to have high
iteration scores; these regulators are all related to the stress
conditions in the training set.

\begin{figure}[htb]
\begin{center}
\includegraphics[scale=0.4]{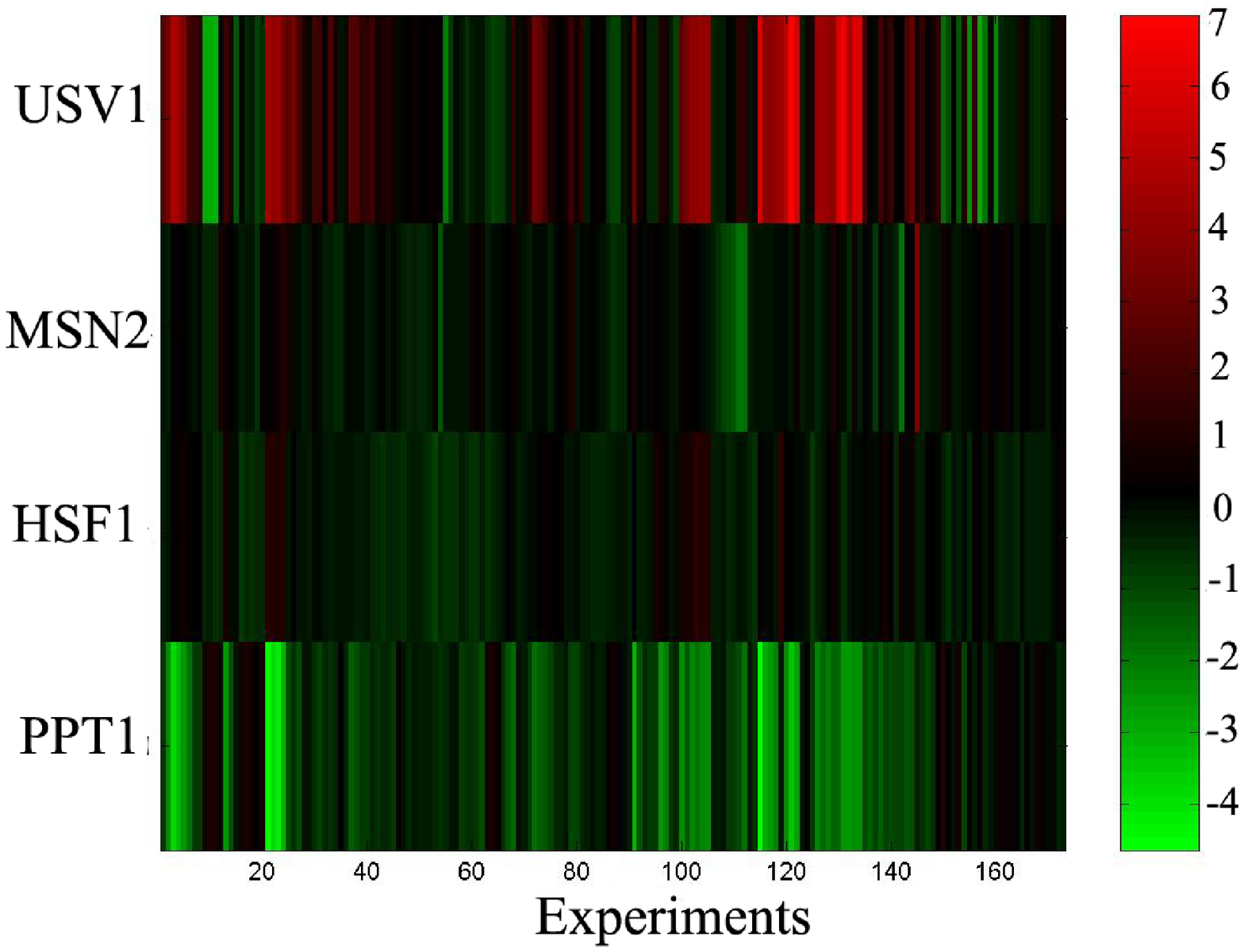}
\caption{\footnotesize {\bf Comparison of expression profiles (173
experiments) of USV1, MSN2, HSF1 and PPT1.}  The mRNA expression
levels of USV1 and PPT1 are informative, with about 50\% and 35\%
of experiments (respectively) showing over 2 fold expression
change over wildtype.  The espression levels for MSN2 and HSF1
fall mostly in the baseline state, with only about 6\% and 5\% of
experiments (respectively) showing at least 2 fold expression
change.  While MSN2 and HSF1 are not identified as high scoring
parents in the learned trees, their binding sites occur as high
scoring motifs.}\label{fig:reg_comp}
\end{center}
\end{figure}

It is interesting to note that while the presence of binding sites
of some very important stress factors like MSN2 and HSF1 (heat
shock factor) are identified as significant features (high motif
iteration score) in the ADT, the mRNA expression levels of these
regulators do not seem to be very predictive. HSF1 does not appear
as a parent and MSN2 gets low abundance and iteration scores as a
parent, despite their importance as heat-shock and general stress
response regulators respectively. Similar results are observed in
the modules of \cite{segal:module}, where HSF1 is not found in any
of the regulation programs and MSN2 is found in 3 of the 50
regulation programs but with low significance. If we compare the
expression profiles of HSF1, MSN2, USV1 and PPT1,
we find that the mRNA levels of MSN2 and HSF1
have small fluctuations (rarely greater than 2 fold change) and
fall mostly within the baseline state, while
the expression levels of USV1 or PPT1 show much larger variation
over many experiments (see Figure \ref{fig:reg_comp}).
It is known that the activity of MSN2 is regulated by TPK1 (a
kinase) via cellular localization. TPK1 is identified as an
important parent in the ADT (AS) and is found associated with the
MSN2 binding site as a motif-parent pair.
Thus in this case, where the activity of a regulator is itself
regulated post-transcriptionally, we see a clear advantage of
using motif data along with mRNA expression data.

{\bf\em USV1 ``in silico'' knockout for heat-shock and osmolarity stress:} \label{sec:insilico}
Using the same training and test microarrays as in the heat-shock/osmolarity
setup, we perform a second study by removing one of the strong regulators, USV1, from the parent
set and retraining the ADT.
We get a
minor but significant increase in test error from 9.3\% to 11\%.

TPK1 in the upregulated state along with the MSN2/MSN4 binding
site is the top scoring feature (IS). TPK1 is also the top scoring
regulator based on abundance.

We also study target genes that change label from correct to
incorrect due to the removal of USV1.  We reason that since these
genes require presence of USV1 in the ADT for correct prediction
of their regulatory response, they are highly dependent on USV1
activity and provide good candidates for downstream targets of
regulatory pathways involving the knocked out parent. We find that
305 target genes change prediction labels. GO annotation
enrichment analysis of these target genes reveal the terms cell
wall organization and biogenesis, heat-shock protein activity,
galactose, acetyl-CoA and chitin metabolism and tRNA processing
and cell-growth. These match many of the terms (namely
transcription factors, nuclear transport, cell wall and transport
sporulation and cAMP pathway, RNA processing, cell cycle, energy,
osmotic stress, protein modification and trafficking, cell
differentiation) enriched by analyzing GO annotations of genes
that changed significantly in a microarray experiment by
\cite{segal:module} with stationary phase induced in a USV1
knockout.

{\bf\em Peroxide, superoxide stress, and pleiotropic response to
diamide:} For the third study, we train on heat-shock,
heat-shock knockouts, over-expression, H$_2$O$_2$ wild-type and
mutant, menadione, DTT experiments, and we test on diamide
experiments. Gasch {\em et al.} \cite{gasch:yeast} consider the
diamide response to be a composite of responses to the experiments
in the training set. We observe a moderate test loss of 16\%,
suggesting that this pleiotropic response is more complex than the
simpler additive responses to heat-shock and osmolarity.

Although USV1, XBP1 and TPK1 are the top three regulators,
we see the emergence of an important parent, YAP1. This factor
appears to be dominant in the ADTs of only those setups that include
redox related stresses, specifically peroxide and superoxide
stresses, in the training sets. It is well documented that YAP1
plays a significant regulatory role in oxidation stresses, and
this role
correlates well with our findings. We hypothesize that USV1 is not
very important for response to diamide based on analysis of the
fourth setup below, and we attribute its presence in the ADT to the
presence of heat-shock experiments in the training set (based on
the first
setup). We thus simulate a knockout by removing USV1 from the
training set and retraining on the training data. Test loss reduces
dramatically from 16\% to 9.2\%, indicating that USV1's presence in
the ADT is detrimental to prediction on diamide response. The ADT
for this setup also shows YAP1 associated with its binding motif
as an important feature (IS).

{\bf\em Redox and starvation response:} In this study, we train on
DTT and diamide stresses and response to nitrogen depletion and
stationary phase induction. We test on diauxic shift and amino
acid starvation experiments. We observe a poor test loss of
27.9\%. This poor prediction accuracy could mean that regulatory
systems active in experiments in the training set and test set are
significantly different. Gasch {et al.}~\cite{gasch:yeast} mention
that the starvation responses are quite different from each other
and significantly more complex than other stresses (DTT, diamide
stress) due to cell-cycle arrest and several secondary effects.

Analysis of the ADT reveals YGL099W (KRE35) as the most abundant
regulator. KRE35 also scores among the top 5 candidates in other
setups involving redox stresses (such as the third setup above).
It could thus be an important regulator for redox related
stresses.

We observe that the poor prediction accuracy correlates with the
absence of USV1 in the ADT, which is otherwise abundant in the
ADTs of all other setups. Since the first three setups show that
USV1 is an important regulator for heat-shock response, we add the
heat-shock experiments to the training set. As expected, on
retraining with this new training set, we get a very significant
improvement in prediction accuracy on the same test set (from
27.9\% to 16\%). This could mean that pathways involved in the
heat-shock response are an important component of the more complex
response to some starvation responses.

\section{Discussion}

We have shown that the GeneClass learning algorithm makes
accurate predictions of gene regulatory response in yeast over
a wide range of experimental conditions.  In particular, in
experiments related to environmental stress response, examination
of the learned GeneClass tree models retrieved important regulators,
motifs, and regulator-motif relationships associated with specific
stress response pathways.  We believe that GeneClass provides a
promising new methodology for integrating expression and regulatory
sequence data to study transcriptional regulation.

One important next step is to use GeneClass to
analyze larger data sets. Since the
Gasch dataset that we used here involves only
environmental stress response experiments, it is likely that many of the
regulatory pathways are not activated and therefore cannot be modeled
by analysis of this dataset alone.
We hope to extend our studies by incorporating additional and more
diverse yeast data sets currently available through resources like
NCBI's Gene Expression Omnibus. At the same time, we
plan to make improvements in the computational efficiency of the
GeneClass software to allow a significant
increase the number of parents so that we can
identify the possible roles of many additional regulators.
In particular, we plan to use using data structures
more appropriate for our pairwise interaction features and
weighted sampling to reduce the size of the memory required for
holding the training data.

A second potential advance would be a more careful treatment of the
raw data.  While the {\it ratio} data (perturbation/wild type) gives a
natural input variable for our analysis, better signal to noise
is likely to be achieved by taking into account the expression levels
separately.  In further work, we plan to use two-color noise modeling
to establish expression-level specific
thresholds and thus allow inclusion of more genes whose up- or
down-regulated states currently fall within the baseline category.  This
improvement will allow more training examples and should enable
us to accurately predict the response of more
subtle target genes.

A third direction for improvement would be to treat parent and child
expression levels as continuous (rather than binary) quantities.
Similarly, the number of motifs
in the regulatory region,
rather than merely their presence/absence,
could be taken into account. While these refinements
could potentially yield more realistic models, it is important that they
be represented in a way that is informative for the learning problem
and avoids overfitting.

\begin{small}
\section*{Acknowledgments}
AK is supported by NSF EEC-00-88001.
CW and MM are partially supported by NSF
ECS-0332479 and NIH GM36277.
CL and CW are supported
by NIH grant LM07276-02, and
CL is supported by an Award in Informatics
from the PhRMA Foundation.
\end{small}

\bibliographystyle{plain}
\bibliography{references}

\begin{thebibliography}{10}

\bibitem{bussemaker:reduce}
H.~J. Bussemaker, H.~Li, and E.~D. Siggia.
\newblock Regulatory element detection using correlation with expression.
\newblock {\em Nature Genetics}, 27:167--171, 2001.

\bibitem{hassle:lin}
P.~D'Haeseleer, X.~Wen, S.~Fuhrman, and R.~Somogyi.
\newblock Linear modeling of m{R}{N}{A} expression levels during {C}{N}{S}
  development and injury.
\newblock {\em Pac. Symp. Biocomp.}, pages 41--52, 1999.

\bibitem{freund:alt}
Y.~Freund and L.~Mason.
\newblock The alternating decision tree learning algorithm.
\newblock {\em Proc. of the Sixteenth International Conf. on Machine Learning},
  pages 124--133, 1999.

\bibitem{gasch:yeast}
A.~P. Gasch, P.~T. Spellman, C.~M. Kao, Orna Carmel-Harel, M.~B. Eisen,
  G.~Storz, D.~Botstein, and P.~O. Brown.
\newblock Genomic expression programs in the response of yeast cells to
  environmental changes.
\newblock {\em Mol. Biol. Cell}, 11:4241--4257, 2000.

\bibitem{hartemink:graphical}
A.~J. Hartemink, D.~K. Gifford, T.~S. Jaakkola, and R.~A. Young.
\newblock Using graphical models and genomic expression data to statistically
  validate models of genetic regulatory networks.
\newblock {\em Pac. Symp. Biocomp.}, pages 422--33, 2001.

\bibitem{hughes:alignace}
J.~D. Hughes, P.~W. Estep, S.~Tavazoie, and G.~M. Church.
\newblock Computational identification of cis-regulatory elements associated
  with groups of functionally related genes in {S}accharomyces cerevisiae.
\newblock {\em J. Mol. Biol.}, 296(5):1205--14, 2000.

\bibitem{ihmels:module}
J.~Ihmels, G.~Friedlander, S.~Bergmann, O.~Sarig, Y.~Ziv, and N.~Barkai.
\newblock Revealing modular organization in the yeast transcriptional network.
\newblock {\em Nature Genetics}, 31:370--377, 2002.

\bibitem{klein:mig1}
C.~J. Klein, L.~Olsson, and J.~Nielsen.
\newblock Glucose control in saccharomyces cerevisiae: the role of {MIG1} in
  metabolic functions.
\newblock {\em Microbiology}, 144:13--24, 1998.

\bibitem{lee:binding}
T.~I. Lee, N.~J. Rinaldi, F.~Robert, D.~T. Odom, Z.~Bar-Joseph, G.~K. Gerber,
  N.~M. Hannett, C.~R. Harbison, C.~M. Thompson, I.~Simon, J.~Zeitlinger, E.~G.
  Jennings, H.~L. Murray, D.~B. Gordon, B.~Ren, J.~J. Wyrick, J.~Tagne, T.~L.
  Volkert, E.~Fraenkel, D.~K. Gifford, and R.~A. Young.
\newblock Transcriptional regulatory networks in {S}accharomyces cerevisiae.
\newblock {\em Science}, 298:799--804, 2002.

\bibitem{middendorf:geneclass}
M.~Middendorf, A.~Kundaje, C.~Wiggins, Y.~Freund, and C.~Leslie.
\newblock Predicting genetic regulatory response using classification.
\newblock http://www.cs.columbia.edu/compbio/geneclass, 2004.

\bibitem{peer:inferring}
D.~Pe'er, A.~Regev, G.~Elidan, and N.~Friedman.
\newblock Inferring subnetworks from perturbed expression profiles.
\newblock {\em Proc. of the Ninth International Conf. on Intelligent Systems
  for Molecular Biology}, pages 215--224, 2001.

\bibitem{peer:minreg}
D.~Pe'er, V.~Regev, and A.~Tanay.
\newblock A fast and robust method to infer and characterize and active
  regulator set for molecular pathways.
\newblock {\em Proc. of the Tenth International Conf. on Intelligent Systems
  for Molecular Biology}, pages 258--267, 2002.

\bibitem{pilpel:motifsyn}
Y.~Pilpel, P.~Sudarsanam, and G.~M. Church.
\newblock Identifying regulatory networks by combinatorial analysis of promoter
  elements.
\newblock {\em Nature Genetics}, 2:153--159, 2001.

\bibitem{schapire:boostexter}
R.~E. Schapire and Y.~Singer.
\newblock Boostexter: A boosting-based system for text categorization.
\newblock {\em Machine Learning}, 39(2/3):135--168, 2000.

\bibitem{Schapire02}
Robert~E. Schapire.
\newblock The boosting approach to machine learning: An overview.
\newblock In {\em MSRI Workshop on Nonlinear Estimation and Classification},
  2002.

\bibitem{SchapireFrBaLe98}
Robert~E. Schapire, Yoav Freund, Peter Bartlett, and Wee~Sun Lee.
\newblock Boosting the margin: {A} new explanation for the effectiveness of
  voting methods.
\newblock {\em The Annals of Statistics}, 26(5):1651--1686, October 1998.

\bibitem{segal:module}
E.~Segal, M.~Shapira, A.~Regev, D.~Pe'er, D.~Botstein, D.~Koller, and
  N.~Friedman.
\newblock Module networks: Identifying regulatory modules and their condition
  specific regulators from gene expression data.
\newblock {\em Nature Genetics}, 34(2):166--176, 2003.

\bibitem{segal:learningmod}
E.~Segal, R.~Yelensky, and D.~Koller.
\newblock Genome-wide discovery of transcriptional modules from {D}{N}{A}
  sequence and gene expression.
\newblock {\em Bioinformatics}, 19:273--282, 2003.

\bibitem{wigender:transfac}
E.~Wingender, X.~Chen, R.~Hehl, H.~Karas, I.~Liebich, V.~Matys, T~. Meinhardt,
  M.~Pr\"uss, I.~Reuter, and F.~Schacherer.
\newblock {TRANSFAC}: an integrated system for gene expression regulation.
\newblock {\em Nucleic Acids Research}, 28:316--319, 2000.

\bibitem{collins:svd}
M.K. Yeung, J.~Tegner, and J.~J. Collins.
\newblock Reverse engineering gene networks using singular value decomposition
  and robust regression.
\newblock {\em Proc. Natl. Acad. Sci. {U}{S}{A}}, 99:6163--8, 2002.

\end{thebibliography}

\end{document}